\documentclass[aps,prl,twocolumn,groupedaddress]{revtex4}

\usepackage{graphicx, epsfig}

\begin{document}

\def\be{\begin{equation}}
\def\ee{\end{equation}}
\def\bd{\begin{displaymath}}
\def\ed{\end{displaymath}}
\def\ba{\begin{eqnarray}}
\def\ea{\end{eqnarray}}

\def\eea{\end{eqnarray}}
\def\bean{\begin{eqnarray*}}
\def\eean{\end{eqnarray*}}

\def\la{\langle} 
\def\ra{\rangle} 
\def\lra{\leftrightarrow} 
\def\bc{\begin{center}} 
\def\ec{\end{center}} 
\def\btab{\begin{tabular}} 
\def\etab{\end{tabular}} 
\def\hl{\hline} 
\def\nono{\nonumber} 
\def\question#1{{{\marginpar{\small \sc #1}}}} 
\def\qq{$ q\bar q $} 
\def\ss{$ s\bar s $} 
\def\nn{$n\bar{n}$} 
\def\uu{$u\bar u$} 
\def\dd{$d\bar d$} 
\def\cc{$c\bar c$} 
 
\oddsidemargin=-4mm 
 
\parindent=0pt

\title{Electroweak production of hybrid mesons in a Flux-Tube simulation of Lattice QCD}
\author{F.E.Close} 
\email{F.Close1@physics.ox.ac.uk} 
\author{J.J.Dudek} 
\email{dudek@thphys.ox.ac.uk}
\affiliation{Department of Physics - Theoretical Physics, University of Oxford,\\
1 Keble Rd., Oxford OX1 3NP, UK}

\begin{abstract} We make the first calculation of the electroweak couplings of hybrid
  mesons to conventional mesons appropriate to photoproduction and to
  the decays of $B$ or $D$ mesons.  $E1$ amplitudes are found to be
  large and may contribute in charge exchange $\gamma p \to n H^+$
  allowing production of (amongst others) the charged $1^{-+}$ exotic
  hybrid off $a_2$ exchange.  Axial hybrid meson photoproduction is
  predicted to be large courtesy of $\pi$ exchange, and its strange
  hybrid counterpart is predicted in $B \to \psi K_H(1^+)$ with $b.r.
  \sim 10^{-4}$.  Higher multipoles, and some implications for hybrid
  charmonium are briefly discussed.

\end{abstract}

\maketitle

An outstanding problem in the Standard Model is how the non-Abelian,
gluon, degrees of freedom behave in the limit of strong QCD.  Lattice
QCD predicts a spectroscopy of glueballs \cite{glue} and hybrid mesons
\cite{hybrids}, but there are no unambiguous signals against which
these predictions can be tested.

A major stumbling block in the case of hybrids is that while
predictions for their masses\cite{hybrids,IP}, hadronic
widths\cite{IK,CP95} and decay channels\cite{IK,CP95,michael2000} are
rather well agreed upon, the literature contains no discussion of
their production rates in electroweak interactions (beyond VMD in one
exotic channel \cite{page}).  Meanwhile a significant plank in the
proposed upgrade of Jefferson Laboratory is its assumed ability to
expose the predicted hybrid mesons in photo and electroproduction.
Also, high statistics studies of meson production in $B$ and $D$
decays are becoming available.

Clearly a calculation of hybrid production, appropriate to such
experiments, in a model based on lattice QCD is urgently called for.
Here we make the first direct calculation of electromagnetic and weak
production of hybrids in such a model \cite{IP}.  We find that the E1
transition amplitudes may be large and accessible in forthcoming
experiments.  Furthermore we predict that $b.r.\left[B \to \psi
  K_H(1^{+})\right] \sim 10^{-4}$, and suggest that evidence for this
may already be present in the enhancement of low momentum $\psi$\cite{bpsidata}.

\subsection*{The model}

Theory\cite{IP,conf} has provided compelling arguments from QCD that
confinement occurs via the formation of a flux tube: a relativistic
object with an infinite number of degrees of freedom.  A standard
approximation\cite{IP,IK,CP95,isgur99} has been to fix the
longitudinal separation $\vec{r} = \vec{r}_Q - \vec{r}_{\bar{Q}}$ and
to solve the flux-tube dynamics in the limit of a thin string with
purely transverse degrees of freedom.  The resulting energies $E(r)$
are then used as adiabatic effective potentials on which the meson
spectroscopies are built.  Ref.\cite{bcs} studied the effect of
relaxing these strict approximations and found that the spectrum of
the conventional and lowest hybrids is robust.  We shall assume the
same is true in this first calculation of electroweak excitation of
hybrid mesons.

In refs.\cite{IP,isgur99,bcs} the flux-tube was discretised into $N+1$ cells, 
and then $N \to \infty$.  Up to $N$ modes may be excited.  We shall focus on the first excited state, with
excitation energy $\omega = \pi /r$. 

The flux tube is dynamic, with degrees of freedom in the two
dimensions transverse to the $Q\bar{Q}$ axis.  The state of the flux
tube can be written in terms of a complete set of transverse
eigenstates $| \vec{y}_1 \ldots \vec{y}_n \ldots \vec{y}_N \rangle $ and the Fourier
mode for the first excited state is

\[ 
\vec{y}_n = \sqrt{\frac{2}{N+1}} \vec{a}_1 \sin \frac{\pi n}{N+1}
\]

In the small oscillation approximation the system becomes
harmonic in $\vec{y} \; (\vec{a})$. The states of the flux-tube are then described
by Gaussians (see eqs.(11,12,13) in ref.\cite{isgur99}). For a pedagogic illustration,
consider the tube to be modeled by a single bead, mass $br$.
 
If the transverse displacement is $\vec{y}$, then conservation of the
position of the centre of mass and of orbital angular momentum about
the centre of mass leads to a mean transverse displacement of the $Q$
and $\bar{Q}$.  If these have masses $m_Q$, then relative to the
centre-of-mass, the position vector of the quark has components in the
longitudinal $\vec{r}$ and transverse $\vec{y}$ directions 
\[
\vec{r}_Q = \left[\frac{1}{2}\vec{r} \; ; \left(\frac{br}{2 m_Q}\right)
  \vec{y}\right] 
\]

The dependence of $\vec{r}_Q$ on $\vec{y}$ enables a quark-current
interaction at $r_Q$ to excite transitions in the $\vec{y}$
oscillator, leading to excitation of the flux-tube.

This is the essential physics behind the excitation of hybrid modes by
current interactions with the quark or antiquark.  Extending to $N$
beads leads to more mathematical detail, but the underlying principles
are the same.  The position vector of the quark becomes \cite{isgur99}

\[ 
\vec{r}_{Q; \bar{Q}} = \vec{R} \pm \frac{1}{2} \vec{r} +\frac{br}{\pi
  m_Q} \sqrt{\frac{2}{N+1}} \vec{a}_1
\] 
 
with $\vec{R}$ the position of the \qq -tube system centre-of-mass. 
 
It has been argued that this dependence $\vec{r}_Q
=f(\vec{r},\vec{y})$ gives significant contributions to static
properties of hadrons, such as charge radii, $\langle r^2
\rangle_{\pi}$ and to the slope of the Isgur-Wise function $\rho(v
\cdot v')$ \cite{isgur99}.  Specifically for $Q\bar{Q}$

\begin{equation} 
r^2_Q = \frac{1}{4}\left[1 + \frac{8b}{\pi^3 m^2_Q} \sum_1^\infty (1/p^3)\right] \langle r^2 \rangle 
\label{r2}  
\end{equation}
 
where the $\sum_1^\infty (1/p^3) \sim 1.2$ arises from the sum over
all modes contributing to zero-point oscillations of the flux-tube.

Isgur\cite{isgur99} showed that these ``transverse excursions" give
huge $\sim 51\%$ corrections in light quark systems where $m_Q = m_d$,
and $\sim 13\%$ corrections in heavy-light $Q\bar{q}$ systems.
Furthermore the $\sum_1^\infty (1/p^3)$ is $\sim 80 \%$ saturated by
its $p=1$ term.  Together, these suggest that the transition
amplitudes to the lowest hybrids ($p = 1$ phonon modes) could be
substantial.  We shall now demonstrate that this can be so, at least
for certain quantum numbers.

The respective amplitudes for conventional $E1$ transitions and the
hybrid excitation come from expanding the incoming plane wave to
leading order in the momentum transfer, thereby enabling the linear
terms in $\vec{q}\cdot \vec{r}_Q$ to break the orthogonality of
initial and final wavefunctions and cause the transition.

By combining with the tensor decomposition of the current-quark
interaction, we may calculate excitation amplitudes to hybrids, and
compare with those for conventional mesons in various multipoles.  We
will give extensive details elsewhere \cite{CD03b}.  In this first
note, we illustrate the principle in electromagnetic interactions and
in what promises to be a prominent heavy flavour decay channel.

A general feature of operators required to excite the lowest hybrid
states (the first flux-tube mode) is the presence of the transverse
position vector $\vec{y}$ to break the orthogonality between the
lowest $Q\bar{Q}$ state and the ``$\vec{y}$-excited" hybrid states.
Hence in photoproduction one accesses $E1$ or (orbitally excited) $M1$
transitions in leading order.  These are $\Delta S=0$, e.g.  $0^{-+}_Q
\to 1^{\pm \pm}_H$ or $1^{--}_Q \to (0,1,2)^{\mp\pm}_H$.  (Note that
states with the ``wrong" charge conjugation will only be accessible
for flavoured mesons, e.g.  in $\gamma p \to H^+ n$, and hence will
have no analogue for $c\bar{c}$ and other $I=0$ states).

Transitions involving spin-flip, $\Delta S =1$, will need a $\vec{\sigma}$
spin operator as well as the above.  Such terms arise as finite size
corrections to the $ \vec{\sigma} \cdot \vec{B}$ magnetic interaction
and also in the spin-orbit interaction $\vec{\sigma} \cdot \vec{p_Q} \times
\vec{E}$, in $J_{em}$.  These are normally non-leading effects at
$O(v/c)^2$ in amplitude and hence much suppressed for heavy flavours.
They are known to give non-negligible contributions to some light
flavour transitions.  However, unlike the leading $\Delta S=0$ terms,
their effects are less well defined (e.g.  binding effects and other
relativistic corrections can play a role at this
order\cite{leyou,lit}).  It is results for the $\Delta S=0$ E1
transitions that are most reliable and on which we primarily focus in
this first evaluation.

\subsection*{E1 excitations} 
 
The familiar $E1$ amplitude between $Q_1\bar{Q}_2$ conventional states (e.g $\gamma \pi
\lra b_1$) is

\begin{equation} 
{\cal M}( \gamma \pi \lra b_1) = \left( \frac{e_1}{m_1} - \frac{e_2}{m_2} \right)   {_b\la r \ra_\pi} |\vec{q}| \frac{\mu}{\sqrt{3}}  \label{e1q} 
\end{equation}

where $_b\la r \ra_\pi$ is the radial wavefunction moment  
$\int_0^\infty r^2 dr R_b(r) r R_\pi(r)$, and $\mu$ is the reduced  
mass of the $Q\bar{Q}$. In line with ref.\cite{isgur99} we use constituent  
masses which subsume contributions from the string. 
 
The analogous amplitude for exciting the $\vec{y}$ oscillator between spin  
singlet states leads to ${\tilde{\cal M}} \equiv {\cal M} \; (\delta_+ - \delta_-)$ 
where 
\begin{equation} 
{\cal M}(\gamma \pi \to a_{1H} ) = \left( \frac{e_1}{m_1} + \frac{e_2}{m_2} \right) {_H\la r \ra_\pi} |\vec{q}| 
\sqrt{\frac{b}{3\pi^3}} \; \delta_{m,+1}  \label{e1h}  
\end{equation}
  
  (where the factors $\delta_{+,-}$ refer to the flux tube $p=1$
  phonon polarisation transverse to the body vector $\vec{r}$, while
  the $\delta_{m,\pm 1}$ refers to the hybrid polarisation in the
  fixed axes $x,y,z$ \cite{IP}).  The transition $\gamma \pi \lra
  a_{1H}$ is seen to vanish when $m_1 \equiv m_2$ and $e_1 = -e_2$ in
  accord with the constraints of $C$ conjugation.  The above formula
  can be immediately taken over to flavoured states where $m_1 \neq
  m_2$.
  
  The parity eigenstates in the flux tube are given in ref.\cite{IP}.
  Following that reference we denote the number of positive or
  negative helicity phonon modes by $\{n_+,n_-\}$, which for our
  present purposes will be $\{1,0\}$ or $\{0,1\}$.  Parity eigenstates
  $\pm$ are then the linear superpositions $\frac{1}{\sqrt 2} \Big(
  |\{1,0\}\rangle \mp |\{0,1\} \rangle \Big) $ such that for $\pi
  \gamma$ $E1$ transitions we have

\[ \langle P = -|\pi \gamma \rangle = 0; \;\;\;\; \langle P = +|\pi \gamma \rangle = \sqrt{2} {\cal M}; \]

This applies immediately to the excitation of the hybrid
$a_{1H}^{\pm}$ in $\gamma \pi^{\pm} \to a_{1H}^{\pm}$ where there is
no spin flip between the spin singlet $\pi$ and $a_{1H}$.  In general
we can write the radiative width $\Gamma (A \to B \gamma)$ as

\[ 
4 \frac{E_B}{m_A}\frac{|\vec{q}|}{(2 J_A + 1)}   
  \sum_{m_J^A} {| {\sqrt{2} \cal{M}}(m_J^A, m_J^B = m_J^A + 1) |}^2 
\]

where the sum is over all possible helicities of the initial meson.
The ratio of widths $\frac{\Gamma_{E1}(a_{1 H}^+ \to \pi^+ \gamma)}
{\Gamma_{E1}(b_{1}^+ \to \pi^+ \gamma)}$ is then

\begin{equation}
\frac{72}{\pi^3} \frac{b}{m_n^2} \left | \frac{_H\la r \ra_\pi}{_b\la r \ra_\pi} \right |^2 \left [ \frac{|\vec{q}_H|^3 \exp \left( - |\vec{q}_H|^2 / {8 \bar \beta_H^2} \right)} {|\vec{q}_b|^3 \exp \left(-|\vec{q}_b|^2 / {8 \bar \beta_b^2} \right) } \right ] 
\end{equation}
 
where the factor in square brackets includes the $q^3$ phase-space and
a ``typical'' form-factor taken from the case of harmonic-oscillator
binding \cite{CDK02}.

Compare the form of this ratio driven by eqs.(\ref{e1q},\ref{e1h}) with the
transverse contribution to the elastic charge radius, eq.(\ref{r2}).  In
the approximation used here, the E1 transitions to the leading states
saturate the dipole sum rule.  This suggests the possibility of
generalising some of our specific results into sum rules relating the
elastic properties of hadrons to the excitation of their hybrid states
\cite{isgur99}.

In the Isgur-Paton adiabatic model\cite{IP} with a variational
harmonic-oscillator solution we obtain $ | _H\la r \ra_\pi / _b\la r
\ra_\pi |^2 \approx 1.0$, so the radial moments do not suppress
hybrids\cite{CD03b}.  We follow ref.\cite{IP} and use the standard
parameters $b=0.18 \mathrm{GeV}^2, m_n=0.33 \mathrm{GeV}$ so that the
prefactor $\frac{72}{\pi^3} \frac{b}{m_n^2} \approx 3.8$ and hence
there is no hybrid suppression from the flux-tube dynamics.

Within our variational solution $\beta_H = 255 \mathrm{MeV}, \beta_b =
281 \mathrm{MeV}, \beta_\pi = 335 \mathrm{MeV}$, so we see the $p=1$
hybrid state being of roughly the same size as the $L=1$ conventional
state. The main uncertainty is the computed size of the
$\pi$\cite{CDK02}.  Assuming that this hybrid has mass $\sim 1.9
\mathrm{GeV}$\cite{hybrids,IP,bcs}, and using the measured width
$\Gamma(b_{1}^+ \to \pi^+ \gamma) = 230 \pm 60 \mathrm{keV}$\cite{PDG}
we predict that

\[
\Gamma(a_{1 H}^+ \to \pi^+ \gamma) = 2.1 \pm 0.9 \mathrm{MeV}.  
\]

where the error allows for the uncertainty in $\beta_{\pi}$\cite{CDK02,CD03b, SWANSON}.  
 
The equivalent $E1$ process for spin triplet $Q\bar{Q}$ states is $(0,1,2)^{+-}_H \lra
\rho \gamma$, where the only difference from the $S=0$ case is the addition of $L,S$
Clebsch-Gordan factors coupling the $Q\bar{Q}$ spin and flux-tube angular momentum to the
total $J$ of the hybrid meson in question.  The matrix element is analogous to
eq.(\ref{e1h}) multiplied by the Clebsch-Gordan $\langle 1+1 ; 1m_\rho | J m_J \rangle$.
We find (for $J=0,1,2$ in this $E1$ limit),
\[
        \Gamma(b_{J H}^+ \to \rho^+ \gamma) = 2.3 \pm 0.8 \mathrm{MeV}. 
\]
where the error reflects the uncertainties in the conventional $E1$ strength and
$\beta_{f_1}$ and where we have taken $m_H = 1.9\mathrm{GeV}$.

\begin{table}
\begin{tabular}{c| c| c| c }
state & & $u\bar{d}$ & $u\bar{s}$  \\
\hline
$^1 S_0$ & $\times 1$ & $\gamma \pi^+ \to a_{1 H}^+$ & $\gamma K^+ \to  K_{1A H}^+$  \\
& & 56 (23) & 43 (23) \\ 
 $^3 S_1$ & $\times \la 11;1m_i | J_H m_H \ra$ & $\gamma \rho^+ \to b_{J H}^+$ & $\gamma K^{*+} \to  K_{J B H}^+$  \\
& & 56 & 43  \\
$^1 P_1$ & $\times \sqrt{\frac{3}{2}} \la 11;1m_i | 1 m_H \ra $ &  $\gamma b_1^+ \to \rho_H^+$ & $\gamma K_{1B}^+ \to  K_H^{*+}$ \\
& & 87 & 68 \\
$^3 P_J$ & $\times \displaystyle{\sum_{m_L,m_S} \la 1 m_L;1 m_S | J m_J \ra} $ & $\gamma a_{J}^+ \to \pi_{J_H H}^+$ & $\gamma K_{JA}^+ \to  K_{J_H H}^+$ \\
& $\cdot \sqrt{\frac{3}{4}} \la 1 m_L+1;1 m_S | J_H m_H \ra^* $ & 87 & 68 \\
\hline
\end{tabular}
\caption{Photon-Meson-Hybrid matrix elements: $\sqrt{2} {\cal M} = \left( \frac{e_1}{m_1} + \frac{e_2}{m_2} \right) \sqrt{2} | \vec{q} | \sqrt{\frac{b}{3 \pi^3}} {_H \la r \ra_i}$ should be multiplied by the Clebsch-Gordan factor in the second column to give the overall matrix element for a positive helicity photon. The numbers quoted in columns three and four are $\sqrt{2} {\cal{M}} /|\vec{q}| \; (10^{-3} \mathrm{GeV}^{-1}) $, evaluated using the results of \cite{IP}, except those
in brackets which use the $\beta$-values of \cite{SWANSON}. 
\label{supertable} 
}
\end{table}

\subsection*{Heavy Flavor Decays} 
 
As discussed after eq.(\ref{r2}), the $|{\cal M}|^2$ for the weak transition $B \to \psi K_H(1^+)$ is
expected to have strength $\sim 13\%$ relative to its ``conventional" counterpart $B \to
\psi K(1^+)$.  Empirically $B^+ \to \psi K(1^+)(1280)$ is the single largest branching
mode in $B^+ \to \psi X$ with $b.r.  = (1.8 \pm 0.5)\times 10^{-3}$ while $B^+ \to \psi
K(1^+)(1400) \leq 0.5 \times 10^{-3}$.  These rates involve both parity conserving (vector) and
violating (axial) contributions and their relative strengths depend on the mixing between
the $^3P_1$ and $^1P_1$ basis states.  These rates would lead one to expect an order of
magnitude $b.r.$ for $B^+ \to \psi K_H(1^+) \geq 10^{-4}$.

Explicit calculation confirms this. (For technical reasons our analysis of heavy-light
dynamics is not identical to the original formulation of \cite{isgur99}.  Details are in
\cite{dudek1}). The transition matrix element has the structure

\[ 
{\cal M} \sim \la K_H | V_{\mu} - A_{\mu} | B \ra f_{\psi}m_{\psi}\epsilon_{\psi}^{*\mu} 
\] 

where $f_{\psi} =0.4 \mathrm{GeV}$\cite{psivalue}.  A non-relativistic expansion of the vector and
axial operators is made for both longitudinal and transverse components and terms linear
in $\vec{y}$ or $\vec{p}_y$ identified.  This is algebraically tedious but in essence
parallels the approach illustrated earlier.  The expectation values of these linear terms
in $\vec{y}$ space generate the transitions to hybrid $K_H(1^+)$; the analogous terms in
$\vec{r}$ space lead to the familiar $K(1^+)$ states.

For $\Delta S =0$ transitions $B \to \psi K_H(1^{\pm})$, 
 
\[
V_{\mu} \sim p_{\vec y}^{\mu} \frac{m_b + m_s}{m_b m_s};  
A_{\mu} \to A_T \sim |\vec{q}| p_{\vec y}^{(T)}/m_bm_s. 
\]
 
Hence the transition to $K_H(1^+)$ is large because the dominant $\la V_{\mu} \ra$
contributes in S-wave; by contrast $K_H(1^-)$ receives its S-wave from the $|\vec{q}| / m_b$-suppressed
$\la A_{\mu} \ra$ while the vector current contributes to P-waves.  Explicit calculation
confirms this where as a function of $m_H = (1.8;2.0;2.2)$GeV we find

\begin{eqnarray*}
b.r [B \to \psi K_H(1^-)] &=& (1.2;0.5;0.2) \times 10^{-5}; \\
b.r [B \to \psi K_H(1^+)] &=& (4.5;2.3;1.0) \times 10^{-4}.  
\end{eqnarray*}

Furthermore we find that $K_H(1^+)$ is dominantly produced with longitudinal polarisation. 
 
While fine details of the model may be questioned, the $O(10^{-4})$ branching ratio to
this hybrid appears robust and accessible to experiment.  It is intriguing therefore that
there is an unexplained enhancement at low $q_{\psi}$, corresponding to high mass $K$
systems, of this magnitude\cite{bpsidata}.

While suggestive, it would be premature to claim this as evidence for hybrid production.
Radial excitations of the $K(1^+)$ are expected in this region, and in the ISGW\cite{isgw}
model, extended to exclusive hadronic decays and assuming standard factorisation arguments
\cite{factor}, we find these to have $b.r.  \sim 10^{-4}$, though slightly less than the hybrid.

Other strange mesons in this mass range are likely to be suppressed due to their high
angular momentum which give powerful orthogonality suppressions at small $q$.  It is the
S-wave character of the hybrid and axial production that drives their significant
production rates.

To test these predictions experimentally, first identify the $\psi$
vertex and reconstruct the B from the decay hadrons and thereby the
invariant mass distribution of the strange system.  Observation of
significant axial strength around 2GeV, produced by parity conserving
S-wave amplitudes at $b.r.  > 10^{-4}$ would prove strong evidence for
the presence of the hybrid meson and warrant further studies of how to
quantify the relative production and mixing of these axial mesons.  In
turn it would underpin our predictions of significant E1 transitions
to such states in photoproduction.
 
There is also the possibility of hybrid Charmonium in $B \to \psi_H
X$. Predicting this involves knowledge of flux-tube formation
dynamics\cite{petrov} which goes beyond the present work.

\subsection*{Conclusions} 
 
We confirm Isgur's conjecture that electromagnetic transitions to hybrids may be
significant.  We find this to be true for certain $E1$ transitions for light flavours in
charge exchange.

Within this model we also anticipate that $gg$ interactions initiate
significant cascades such as $\psi_H \to \psi \eta(\eta \prime)$ and
the diffractive transition $\gamma N \to 2^{+-}_H N$; these currents
will disturb the flux tube by direct analogy with the electromagnetic
transitions discussed here.

These results promise an active programme of future research at an upgraded Jefferson
Laboratory and at CLEO-c.  They also encourage mining existing data on $B$ decays and
inclusion in future plans for heavy flavor decays.  In particular there is the intriguing
observation of an as yet unexplained enhancement in $B \to \psi X$ in the kinematic region
where $K_H$ is expected, and with a strength compatible with that predicted for $K_H =
1^+$.  We urge further investigation of this, and the other channels identified in this
note.

We shall give a detailed discussion elsewhere\cite{CD03b}.

\bc 
{\bf Acknowledgments} 
\ec 
 
This work is supported, in part, by grants from the Particle Physics and Astronomy Research Council,  and the EU-TMR program ``Euridice'', HPRN-CT-2002-00311. We thank P.J.S. Watson for discussions on the dynamics of flux tubes.

\end{document}